\documentstyle[11pt,twoside,fancyhea,cmp]{article}
\title{\Large\bf
CORRELATION FUNCTIONS OF QUENCHED AND ANNEALED ISING SYSTEMS
   }
\author{
   {\sc R.~R.~Levitskii, S.~I.~Sorokov, R.~O.~Sokolovskii}
   \\[1.5ex]
   \it Institute for Condensed Matter Physics
   \\
   \it of the Ukrainian National Academy of Sciences
   \\
   \it 1~Svientsitski St., UA--290011 Lviv, Ukraine
}
\date{Received March 2, 1995}

\newcommand{\bee}{\begin{eqnarray}}
\newcommand{\eee}{\end{eqnarray}}
\newcommand{\be}{\begin{equation}}
\newcommand{\ee}{\end{equation}}
\newcommand{\braced}[1]{{\left\{#1\right\}}}
\newcommand{\ptsed}[1]{\left(#1\right)}
\newcommand{\avX}[1]{\langle #1 \rangle_x}
\newcommand{\AV}[1]{\langle #1 \rangle_H}
\newcommand{\mref}[1]{(\ref{#1})}
\newcommand{\Hams}{H_x\ptsed{\braced{S}}}
\newcommand{\Ham}{H\ptsed{\braced{S,X}}}
\newcommand{\Sp}{{\rm Sp}}
\newcommand{\as}{_\alpha}
\newcommand{\ab}{_{\alpha\beta}}

\newcommand{\ia}{_{i\alpha}}

\newcommand{\jb}{_{j\beta}}

\newcommand{\iajb}{_{i\alpha,j\beta}}

\newcommand{\fr}{\, _r\bar{\varphi}}
\newcommand{\ps}{\, _r\bar{\psi}}
\newcommand{\kappat}{\bar{\kappa}}
\newcommand{\kt}[1]{\, _{#1}\kappat}
\newcommand{\phidif}[2]{\, _{#1}\hat{\bar{\varphi}}^{(1)}_{#2}}
\newcommand{\largeKappa}{\mbox{\boldmath$\kappa$}}	
\newcommand{\largePhi}{\mbox{\boldmath$\varphi$}}	
\newcommand{\largeS}{{\bf S}}	
\newcommand{\largeJ}{{\bf J}}
\newcommand{\largeGamma}{{\bf \Gamma}}
\newcommand{\ie}{{\it i.e. }}
\newcommand{\ang}[1]{\langle #1\rangle}
\newcommand{\pdif}[2]{\frac{\partial #1}{\partial #2}}
\newcommand{\fdif}[2]{\frac{\delta #1}{\delta #2}}
\newlength{\trialen}
\newcommand{\tria}[5]{
\unitlength=\trialen
\linethickness{0.4pt}
\raisebox{-10\trialen}{	
\begin{picture}(22.00,20.00)
\put(1,18){\line(1,0){20}}
\put(1,18){\line(2,-3){10}}
\put(21,18){\line(-2,-3){10}}
\put(11,12){\makebox(0,0)[cb]{$ #1 $}}
\put(0,18){\makebox(1,0)[rb]{$ #2 $}}	
\put(21,18){\makebox(3,0)[lb]{$ #3 $}}	
\put(11,2){\makebox(0,0)[ct]{$ #4 $}}
\put(13,3){\makebox(3,0)[lc]{$ #5 $}}
\end{picture}	}
}

\newcommand{\upp}[1]{
\raisebox{7\trialen}{$#1$}
}
\newcommand{\sect}[1]{\section{#1}\setcounter{equation}{0}}
\renewcommand{\theequation}{\arabic{section}.\arabic{equation}}

\begin{document}\maketitle
\begin{abstract}
{\small
Spin correlation functions (up to the 3-site one) of disordered
Ising model with the nearest neighbour interaction are calculated and
investigated within a two-site cluster approximation for both quenched
and annealed cases. The approach yields the exact results for
the one-dimensional system. The long-range interaction is taken
into account in the mean field approximation.}
\end{abstract}

\sect{Introduction}
During recent years the statistical  theory  of  disordered
spin models is an object of great interest due  to  their  various
properties and applications. In this paper we will  consider  an
alloy of Ising systems (AIS) that can describe the processes  of
(magnetic or ferroelectric) ordering in alloys of magnets,  magnets
with nonmagnetic admixtures, solid solutions of  ferroelectrics
etc. This model is considered usually in two limiting  cases,
namely the case of annealed  system  (equilibrium  type  of
disorder) and the case of quenched  system  (nonequilibrium
disorder). There is no common opinion on the question, what
type of disorder is realized in partially deuterrated ferroelectrics,
e.g. $Cs(D_xH_{1-x})_2PO_4$, $K(D_xH_{1-x})_2PO_4$. This situation
requires both annealed and quenched cases to be theoretically studied and
compared with experimental data.

The exact results for both limits are known  at the
one-dimensional case \cite{Balagurov}. Effect of dilution (noninteracting
admixture) on the two- and three-dimensional Ising model (quenched case)
have been studied by Monte-Carlo method \cite{Ching,Stoll}. AIS can  also  be
used as a test of approximative methods  of  statistical  mechanics.
It is known that the mean field approximation (MFA) yields
quite satisfactory results for the Ising model.  However  it  is
not able to reproduce some essential properties of AIS \cite{Vaks} (such
as percolation in diluted system and  some  differences  between
the properties of annealed and quenched systems).

In  spite  of
great attention to AIS and great advance  in  its  investigation
(see review in \cite{SatoDilute,prepr1}) the problem of  calculation
of  AIS's correlation functions (CFs) for the case $D\ge 2$  (within  some
acceptable approximation) is not solved up to now. In  this  paper
we suggest the solution of this problem for both quenched and annealed
systems generalizing the approach of \cite{Yukhn} on disordered model.
The short-range interactions are taken into account  within  the
two-site cluster approximation (TCA). In the appendix the long-range
interactions are also taken into consideration within MFA.

\sect{Formulation of the problem}
We shall consider an Ising mixture with the Hamiltonian
${\cal H}(\braced{S,X})$
\be
-\beta {\cal H}(\braced{S,X})\equiv \Ham
=\Hams+H(\braced{X});\qquad \beta=1/k_BT\label{Hamilt}\;,
\ee
where
\be
\Hams=\sum_i\kappa_i S_i +\frac{1}{2}\sum_{ij} K_{ij} S_i S_j
\ee
describes a quasispin subsystem on the sites of a simple lattice
($i,j=1\cdots N$) with
the pair exchange interaction $K_{ij}$ in the site dependent field
$\kappa_i$. The set of spin variables $\braced{S_i}$ represents a state of
the spin subsystem ($S_i=\pm 1$).

The other part of the Hamiltonian
\be
H(\braced{X})=\sum_i\mu_i+\frac{1}{2}\sum_{ij} V_{ij}
\ee
includes "nonexchange" pair interaction $V_{ij}$ and field
$\mu_i$. Each site contains a spin of a certain sort and the sort
configuration is described by the set of variables $\braced{X\ia}$
($\alpha=1\cdots \Omega$, $\Omega$ is a number of sorts):
$X\ia=1$ if the site $i$ is occupied by the spin of sort $\alpha$,
otherwise $X\ia=0$. Interactions $\kappa_i$, $K_{ij}$, $\mu_i$,
$V_{ij}$ depend on the sort configuration
\bee
&&\mu_i = \sum\as \mu\ia X\ia ; \qquad
V_{ij} = \sum\ab V\iajb X\ia X\jb 	\;;\nonumber\\
&&\kappa_i = \sum\as \kappa\ia X\ia ; \qquad
K_{ij} = \sum\ab K\iajb X\ia X\jb	\;.
\eee
It should be noted that parameters of the Hamiltonian contain inverse
temperature $\beta$.
We suppose the pair interactions to be short-range
\be
 V\iajb = \beta V\ab\pi_{ij} ;\quad \beta K\iajb = K\ab\pi_{ij};\quad
 \pi_{ij} = \left\{
\begin{array}{ll}
        1, & $if $ i\in\pi_j\\
        0, & $in opposite case$
\end{array}
             \right. 	\;,
\ee
where $\pi_i$ denotes the set of the nearest neighbours of the site $i$
(the first coordination sphere). In an appendix we will also take into
account the long-range interactions in MFA.

The model is considered at the cases of annealed system and of
quenched one. At the first case an equilibrium sort configuration is
realized and the system is described by the density
matrix
\newcommand{\F}{{\cal F}}
\newcommand{\Z}{{\cal Z}}
\be
\rho(\braced{S,X})=\Z^{-1} \exp \Ham; \qquad \Z=\Sp_\braced{S,X}\exp \Ham
\ee
as well as by the generating function
\be
\F(\braced{\kappa,\mu})=\ln \Z=-\beta {\cal G},
\ee
where ${\cal G}$ is the grand thermodynamic potential of the system.

The generating function allows to calculate the correlation
functions (CFs) of the system
\bee
&&\ang{
S_{i_1\alpha_1}\cdots S_{i_n\alpha_n}X_{j_1\beta_1}\cdots X_{j_l\beta_l}
}^c		=\nonumber\\
&&=\fdif{}{\kappa_{i_1\alpha_1}}\cdots\fdif{}{\kappa_{i_n\alpha_n}}
\fdif{}{\mu_{j_1\beta_1}}\cdots\fdif{}{\mu_{j_l\beta_l}}
\F(\braced{\kappa,\mu})	\;,
\eee
where $S\ia=S_iX\ia$, 		
the superscript $^c$ means cumulant averaging and
\be
\ang{(\cdots)}=\Sp_\braced{S,X} \left[ \rho(\braced{S,X})(\cdots) \right]\;.
\ee
The chemical potentials $\mu\ia$ have to be found from the equations
\be
\ang{X\ia}=c_\alpha \label{mu equation}	\;,
\ee
where $c_\alpha$ is a concentration of the sort $\alpha$ spins.

     At the case of quenched type of disorder the sort configuration is fixed
and independent of temperature,
therefore thermodynamic averaging implies only a trace over spin degrees of
freedom
\be
\AV{(\cdots)} = \Sp_\braced{S} \left[ \rho_x(\braced{S})(\cdots) \right] ;\qquad
\rho_x(\braced{S}) = Z_x^{-1}\exp \Hams	\;;
\ee\be
Z_x = \Sp_\braced{S} \exp(\Hams)	\;.
\ee
In order to obtain observable quantities one must perform also an averaging
over sort configurations
\be
\avX{(\cdots)}  = \Sp_\braced{X} \left[ \rho(\braced{X})(\cdots) \right]\;,
\ee
where the distribution $\rho(\braced{X})$ is determined by the conditions of
system's freezing. Our approximation
is sensitive to the following moments of this distribution:
\be
    \avX{X\ia X\jb} = w\ab\, \;
  (j \in \pi_i) ; \qquad \avX{X\ia} = c\as  \equiv \sum_\beta w\ab\;.
\ee
Free energy of a quenched system is defined as follows
\be
F = -\beta^{-1}\avX{F_x}-\beta^{-1} \avX{H(\braced{X})} ;\qquad
\avX{F_x}=\avX{\ln \Sp_\braced{S}\exp[\Hams]}\;.
\label{freeen}
\ee
Function $\avX{F_x}=\avX{F_x}\ptsed{\braced{\kappa}}$ that is the cumulant generating function in this case:
\be
m^{(n)}_{i_1\alpha_1\cdots i_n\alpha_n}=
  \avX{\AV{S_{i_1\alpha_1}\cdots S_{i_n\alpha_n} }^c} =
\fdif{ }{\kappa_{i_1\alpha_1}}\cdots\fdif{ }{\kappa_{i_n\alpha_n}}
\avX{F_x}\ptsed{\braced{\kappa}} \;.
         \label{(2.12)}
\ee
\sect{Quenched case}
First we carry out a cluster expansion of the generating
function $\avX{F_x}$. For this purpose we write the Hamiltonian
$\Hams$ in the form
\be
\Hams = \sum_i H_i + \sum_{(ij)}U_{ij} \;,
        \label{(3.1)}
\ee
where
\be
H_i = \kappat_i S_i ;\qquad \kappat_i = \kappa_i + \sum_{r\in\pi_i}
  \fr_i;\qquad    \fr_i = \sum\as X\ia \fr\ia \label{def_of_kappat} \; ;
\ee\be
U_{ij} = -\ _j\bar{\varphi}_i S_i -\ _i\bar{\varphi}_j S_j + K_{ij} S_i S_j\;.
\ee

We introduce here the parameters $\fr_i$, which play a role of the effective
field acting on the spin $S_i$ from the nearest neighbour
at the site $r$. Summation $\sum_{(ij)}$ in \mref{(3.1)} spans pairs of
the nearest neighbour sites. We can write function $F_x$ in the form
\be
         F_x = \sum_i F_i  + \ln Q  \;.
\ee
Here we use the notations
\[
 F_i = \ln Z_i ;\qquad Z_i = \Sp_{S_i}\exp H_i	\;;
\]\be
 Q = \ang{\exp\sum_{(ij)}U_{ij} }_{\rho_0};\qquad
     \rho_0 = \prod_i \rho_i ;\qquad \rho_i = Z_i^{-1}\exp(H_i) \;.
\ee
The first term ${\cal K} _1$ of a cluster expansion \cite{prepr1,Kubo}
of $\ln Q$ has the form
\be
 \ln Q = \ln\ang{\exp\sum_{(ij)}U_{ij}}_{\rho_0} \approx {\cal K}_1 =
 \sum_{(ij)}\ln\ang{\exp U_{ij}}_{\rho_0} =
 - z\sum_i F_i + \sum_{(ij)} F_{ij}	\;,
\ee
where $z=\sum_j\pi_{ij}$ is a number of the nearest neighbour sites and
\bee
&&F_{ij} = \ln Z_{ij};\qquad Z_{ij} = \Sp_{S_iS_j} \exp H_{ij};\nonumber\\
&&H_{ij} =  \kt{i}_j S_i  +  \kt{j}_i S_j+ K_{ij}S_i S_j;\qquad
\kt{r}_i = \kappat_i - \fr_i	\;.
\eee
So the generating function $\avX{F_x}$ restricted by the first term
of the cluster expansion is obtained in the form
\be
  \avX{F_x} = -z'\sum_i\avX{F_i} + \sum_{(ij)}\avX{F_{ij}}\quad (z'=z-1).
\ee
Parameters $\fr\ia$ are found due to the minimization of the free energy
\be
   \pdif{\avX{F_x}}{\fr\ia} = 0       \label{(3.8)}	\;.
\ee
It gives the following self-consistency equations for the average value of
spin $m^{(1)}\ia=\avX{\AV{S\ia}}$
\bee
m^{(1)}\ia&=&\avX{\ang{S\ia}_{\rho_i}}\label{3.9a}	\;,\\
\avX{\ang{S\ia}_{\rho_i}}&=&\avX{\ang{S\ia}_{\rho_{ij}}}\label{3.9b}	\;,
\eee
where $j \in \pi_i$ and
\be
\rho_{ij}=\frac{\exp H_{ij}}{Z_{ij}} \;.
\ee
Relations \mref{3.9a}, \mref{3.9b} can be written in the form:
\bee
m^{(1)}\ia&=&\avX{F^{(1)}\ia}\label{other3.9a}\;,\\
\avX{F^{(1)}\ia}&=&\avX{F^{(10)}_{i\alpha,r}}\label{other3.9b}\;.
\eee
\begin{figure}[t]
\unitlength=1mm
\centerline{
 \begin{picture}(120,98)
 \put(0,98){\special{em:graph m.pcx}}
 \end{picture}
}
\caption{
Order parameters of pure and dilute ($w_{11}=c_1^2$)
systems on plane square lattice ($z=4$).
}
\label{diferent M}
\end{figure}
Here the notations
\be
\avX{F^{(n)}_{i\alpha_1\cdots\alpha_n}}=
\pdif{}{\kappa_{i\alpha_1}}\cdots\pdif{}{\kappa_{i\alpha_n}} \avX{F_i}=
\avX{\ang{S_{i\alpha_1}\cdots S_{i\alpha_n}}^c_{\rho_i}}	\;;
\ee\bee
\avX{F^{(nl)}_{i\alpha_1\cdots\alpha_n,j\beta_1\cdots\beta_l}}&=&
\pdif{}{\kappa_{i\alpha_1}}\cdots\pdif{}{\kappa_{i\alpha_n}}
\pdif{}{\kappa_{j\beta_1}}\cdots\pdif{}{\kappa_{j\beta_l}} \avX{F_{ij}}=\nonumber\\
&=&\avX{\ang{S_{i\alpha_1}\cdots S_{i\alpha_n}
   S_{j\beta_1}\cdots S_{j\beta_l}}^c_{\rho_{ij}}}
\eee
are introduced. Below we give the explicit expressions  for  the
quantities $\avX{F_i}$, $\avX{F_{ij}}$ and their derivatives which will be
used later:
\bee
&& \avX{F_i}=\sum\as c\as F\ia;\qquad F\ia=\ln 2\cosh \kappat\ia	\;;\nonumber\\
&& \avX{F^{(n)}_{i\alpha_1\cdots\alpha_n}}=\delta_{\alpha_1\alpha_2}
\cdots \delta_{\alpha_{n-1}\alpha_n}c_{\alpha_1}F^{(n)}_{i\alpha_1}	\;;\nonumber\\
&& F^{(1)}\ia=\tanh \kappat\ia;\qquad
F^{(2)}\ia=1-\tanh^2 \kappat\ia;\qquad
F^{(3)}\ia=-2F^{(1)}\ia F^{(2)}\ia   \;;\nonumber\\
&&\avX{F_{ij}}=\sum\ab w\ab F\iajb; \qquad F\iajb=\ln 2 e^{K\ab}L\iajb\;;\nonumber\\
&&L\iajb=\cosh(\kt{j}\ia+\kt{i}\jb)+a\ab\cosh(\kt{j}\ia-\kt{i}\jb);
\quad a\ab=e^{-2K\ab}  \;;\nonumber\\
&&\avX{F^{(n0)}_{i\alpha_1\cdots\alpha_n,j}}=
\delta_{\alpha_1\alpha_2}\cdots \delta_{\alpha_{n-1}\alpha_n}
\sum_\beta w_{\alpha_1\beta}F^{(n0)}_{i\alpha_1,j\beta}   \;;\nonumber\\
&&F\iajb^{(10)}=
\ptsed{\sinh(\kt{j}\ia+\kt{i}\jb)+a\ab\sinh(\kt{j}\ia-\kt{i}\jb)}/
{L\iajb}   \;;\nonumber\\
&&F\iajb^{(20)}=1-\ptsed{F\iajb^{(10)}}^2=
\ptsed{1+a\ab^2+2a\ab \sinh(2\kt{i}\jb)}/{L\iajb^2}  \;;\nonumber\\
&&F\iajb^{(30)}=-2F\iajb^{(10)}F\iajb^{(20)}    \;;\nonumber\\
&&\avX{F^{(nl)}_{i\alpha_1\cdots\alpha_n,j\beta_1\cdots\beta_l}}=
\delta_{\alpha_1\alpha_2}\cdots \delta_{\alpha_{n-1}\alpha_n}
\delta_{\beta_1\beta_2}\cdots \delta_{\beta_{l-1}\beta_l}
w_{\alpha_1\beta_1} F^{(nl)}_{i\alpha_1,j\beta_1}; \nonumber\\
&&F\iajb^{(11)}=\frac{1-a\ab^2}{L\iajb^2} ; \qquad
F\iajb^{(21)}=-2F\iajb^{(10)}F\iajb^{(11)}	\;.
\eee
The expression \mref{3.9b} contains $Nz\Omega$ equations for the same
number of variables $\fr\ia$.  At the case of uniform field
($\kappa\ia\longrightarrow\kappa\as$) parameters $\fr\ia$ lose their site
dependence and \mref{3.9b} reduces to $\Omega$ equations for the same
number of fields and we obtain a well-known result of Bethe approximation
and cluster variation method for average value of spin
\be
m^{(1)}\as=c\as\tanh \kappat\as=\sum_\beta w\ab
\frac{\sinh(\kappat'\as+\kappat'_\beta)+a\ab\sinh(\kappat'\as-\kappat'_\beta)}
{\cosh(\kappat'\as+\kappat'_\beta)+a\ab\cosh(\kappat'\as-\kappat'_\beta)} \;,
\ee\be
\kappat\as=\kappa\as+z\bar{\varphi}\as;
\qquad\kappat'\as=\kappa\as+z'\bar{\varphi}\as	\;.
\ee
We can obtain CFs of any order with differentiation of \mref{3.9a},
\mref{3.9b} and it is an advantage of presented approach.

It follows from \mref{other3.9a} that
\be
m^{(2)}\iajb=\fdif{m^{(1)}\ia}{\kappa\jb}=
\pdif{\avX{F^{(1)}\ia}}{\kappat_{i\gamma}}\fdif{\kappat_{i\gamma}}{\kappa\jb}=
\avX{F^{(2)}_{i\alpha\gamma}}\kappat^{(1)}_{i\gamma,j\beta} \;.
\label{m2}
\ee
Hereafter the summation over sort indices (Greek letters) is not
written explicitly so the sum over $\gamma$ symbol is omitted in \mref{m2}.
We have just used the notation
\be
\label{defofhatkap}
\bar{\kappa}^{(1)}\iajb=
\fdif{\bar{\kappa}\ia}{\kappa\jb}=
\delta_{ij}\delta\ab+\sum_{r\in\pi_i}\,_r\bar{\varphi}^{(1)}\iajb;
\qquad \bar{\varphi}^{(1)}\iajb= \fdif{\,_r\bar{\varphi}\ia}{\kappa\jb}
\ee
(see a definition of $\kappat_i$ \mref{def_of_kappat}).
It is convenient to write the relation \mref{m2} in the matrix form
\be
\hat{m}^{(2)}_{ij}=\hat{F}^{(2)}_i\hat{\kappat}^{(1)}_{ij}
\label{mefk}	\;,
\ee
where the matrix notation is introduced:
\be
\ptsed{\hat{m}^{(2)}_{ij}}\ab=m^{(2)}\iajb \;,
\ptsed{\hat{F}^{(2)}_i}\ab=\avX{ F^{(2)}_{i\alpha\beta} } \;,
\ptsed{\hat{\kappat}^{(1)}_{ij}}\ab=\kappat^{(1)}\iajb	\;.
\ee
In order to calculate the quantities $\fr\iajb^{(1)}$ it is necessary to use
the relation \mref{other3.9b}. Differentiating it with respect to
$\kappa\jb$ we obtain:
\be
\hat{F}^{(2)}_i\kappat^{(1)}_{ij}=
\hat{F}^{(20)}_{ir}\,_r\hat{\kappat}^{(1)}_{ij}+
\hat{F}^{(11)}_{ir}\,_i\hat{\kappat}^{(1)}_{rj}
\label{fkefk}	\;,
\ee
where
\be
\ptsed{\hat{F}^{(20)}_{ir}}\ab=\avX{ F^{(20)}_{i\alpha\beta,r} };\qquad
\ptsed{\hat{F}^{(11)}_{ir}}\ab=\avX{ F^{(11)}_{i\alpha,r\beta} }\;;
\ee\be
_r\hat{\kappat}^{(1)}_{ij}=\hat{\kappat}^{(1)}_{ij}-\phidif{r}{ij} ;\qquad
\ptsed{\phidif{r}{ij}}\ab=\fr^{(1)}\iajb
\label{kf}\;.
\ee

\begin{figure}
\unitlength=1mm
\centerline{
 \begin{picture}(110,160)
 \put(0,160){\special{em:graph phases1g.pcx}}
 \end{picture}
}
\newcounter{phasecounter}
\caption{
Phase diagrams for dilute system on plane square lattice ($z=4$) at
different "nonexchange" interaction $\tilde{V}_{11}$).  Annealed:
\setcounter{phasecounter}{1}\Roman{phasecounter} -- paraphase,
\setcounter{phasecounter}{2}\Roman{phasecounter} -- spin ordered phase,
\setcounter{phasecounter}{3}\Roman{phasecounter} -- segregated alloy.
}
\label{PhaseDiags1}
\end{figure}
Taking into account \mref{kf}, the equation \mref{fkefk} can be rewritten
in the following form:
\be
\hat{F}^{(20)}_{ir}\,_r\hat{\bar{\varphi}}_{ij}^{(1)}+
\hat{F}^{(11)}_{ir}\,_i\hat{\bar{\varphi}}_{rj}^{(1)} 
=\ptsed{\hat{F}^{(20)}_{ir}-\hat{F}^{(2)}_i}\hat{\kappat}^{(1)}_{ij}+
\hat{F}^{(11)}_{ir}\hat{\kappat}_{rj}^{(1)}
\label{fphi1}	\;.
\ee
Replacing the indices $i\leftrightarrow r$ in \mref{fphi1} we get the second
equation
\be
\hat{F}^{(11)}_{ri}\,_r\hat{\bar{\varphi}}_{ij}^{(1)}+
\hat{F}^{(20)}_{ri}\,_i\hat{\bar{\varphi}}_{rj}^{(1)} 
=\hat{F}^{(11)}_{ri}\hat{\kappat}_{ij}^{(1)}
+\ptsed{\hat{F}^{(20)}_{ri}-\hat{F}^{(2)}_r}\hat{\kappat}^{(1)}_{rj}
\label{fphi2}	\;,
\ee
which forms together with \mref{fphi1} the set of equations for the
quantities $\phidif{r}{ij}$, $\phidif{i}{rj}$. Excluding $\phidif{i}{rj}$
from this set and expressing the $\hat{\kappat}^{(1)}_{ij}$ via
$m^{(2)}_{ij}$ (see \mref{mefk}) we have
\be
_r\hat{\bar{\varphi}}_{ij}^{(1)}=
\left[(\hat{F}^{(2)}_i)^{-1}+\hat{E}_{ii}\right]\hat{m}^{(2)}_{ij}+
\hat{G}_{ir}\hat{m}^{(2)}_{rj}
\label{4.16}	\;,
\ee
where
\bee
\label{4.17}
\hat{E}_{ii}&=&\left(\hat{F}^{(11)}_{ir}\ptsed{\hat{F}^{(20)}_{ri}}^{-1}\hat{F}^{(11)}_{ri}-\hat{F}^{(20)}_{ir}\right)^{-1}\;,\nonumber\\
\hat{G}_{ir}&=&\left(\hat{F}^{(20)}_{ri}\ptsed{\hat{F}^{(11)}_{ir}}^{-1}\hat{F}^{(20)}_{ir}-\hat{F}^{(11)}_{ri}\right)^{-1}\;.
\eee
Later on we sum the relation \mref{4.16} over $r\in\pi_i$ and take into
account \mref{mefk}, \mref{defofhatkap}. It results in the following Ornstein-Zernike-type
equation for the pair CF $m^{(2)}_{ij}$
\be
\hat{U}_{ii}\hat{m}^{(2)}_{ij}=\delta_{ij}+\sum_{r\in\pi_i}\hat{G}_{ir}\hat{m}^{(2)}_{rj}
\label{OZeq}	\;,
\ee
where
\be
\hat{U}_{ii}= -z'\ptsed{\hat{F}^{(2)}_i}^{-1}-\sum_{r\in\pi_i}\hat{E}_{ii}
\label{4.19}	\;.
\ee
In the case of uniform field the matrices $\hat{F}$ are independent of
site indices, the CF $\hat{m}^{(2)}_{ij}$ depends on the difference of the
sites' 1, 2 coordinates and we can solve the equation \mref{OZeq} carrying
out the Fourier transformation
\be
\hat{m}^{(2)}_{ij}=\hat{m}^{(2)}(\vec{r}_j-\vec{r}_i)=
{V_e\over (2\pi)^D}\int d\vec{q}\;
e^{-i\vec{q}(\vec{r}_j-\vec{r}_i)}
\hat{m}^{(2)}(\vec{q}) \;,   \label{3.20}
\ee
\begin{figure}
\unitlength=1mm
\centerline{
 \begin{picture}(120,120)
 \put(0,120){\special{em:graph phases2.pcx}}
 \end{picture}
}
\caption{
MFA results for the phase diagram of the system. $J\ab=\sum_j J\iajb$,
$\tilde{I}_{11}=\sum_j(I_{i1,j1}-2I_{i1,j2}+I_{i2,j2})$ (see the Appendix).
}
\label{PhaseDiags2}
\end{figure}
$V_e$ is a volume of the elementary cell, $D$ being the dimension of
the lattice (space).
The solution has the form
\bee
&&\label{4.21}\left(\hat{m}^{(2)}(\vec{q})\right)^{-1}=-z'(\hat{F}^{(2)})^{-1}+ z\left(\hat{F}^{(20)}+\hat{F}^{(11)}\right)^{-1}+\nonumber\\
&&+\left(\pi(\vec{0})-\pi(\vec{q})\right)\left(\hat{F}^{(20)}\ptsed{\hat{F}^{(11)}}^{-1}\hat{F}^{(20)}-\hat{F}^{(11)}\right)^{-1}\;,
\eee
where
\be
\pi (\vec{q})=\sum_{j\in\pi_i}e^{i\vec{q}(\vec{r}_j-\vec{r}_i)} ; \qquad
\pi (\vec{0})= z.
\ee
For the hypercubic lattices one has
\be
\pi (\vec{q})=2\sum^D_{l=1}\cos (\vec{q}_l a)	\;,
\ee
where $a$ is the lattice spacing.

The expression \mref{4.21} in the limit $c_1=1$ yields the TCA result for
the ideal (one-sort) system \cite{Yukhn}. Taking $\vec{q}=0$ in \mref{4.21}
one gets the earlier result \cite{Vaks,prepr1}. The known exact formula for
the one-dimensional system in zero external field \cite{Balagurov}
also follows from \mref{4.21}.

   We will obtain the expression for the 3-site CF
$m^{(3)}_{1\alpha,2\beta,3\gamma}$ differentiating
equation \mref{OZeq} which can be written in the form
\be
U_{1\alpha,1\delta} m^{(2)}_{1\delta,2\beta}=\delta_{12}\delta\ab+\sum_{r\in\pi_1}G_{1\alpha,r\delta}m^{(2)}_{r\delta,2\beta}
\ee
(here free site indices are denoted with numbers and
summation over repeated sort indices is implied).
The differentiation with respect to $\kappa_{3\gamma}$ yields
\be
\label{5.2}
\dot{U}_{1\alpha,1\delta,3\gamma} m^{(2)}_{1\delta,2\beta}+ U_{1\alpha,1\delta}m^{(3)}_{1\delta,2\beta,3\gamma}=
\sum_{r\in\pi_1}\left(\dot{G}_{1\alpha,r\delta,3\gamma}m^{(2)}_{r\delta,2\beta}+G_{1\alpha,r\delta}m^{(3)}_{r\delta,2\beta,3\gamma}\right)
\;,
\ee
where the point denotes the derivatives of corresponding quantities:
\bee
\label{5.3}&&\dot{U}_{1\alpha,1\delta,3\gamma}= \fdif{U_{1\alpha,1\delta}}{\kappa_{3\gamma}}
\equiv\tria{\dot{U}}{_{1\alpha}}{_{1\delta}}{_{3\gamma}}{}	\;,\nonumber\\
&&\dot{G}_{1\alpha,r\delta,3\gamma}= \fdif{G_{1\alpha,r\delta}}{\kappa_{3\gamma}}
\equiv\tria{\dot{G}}{_{1\alpha}}{_{r\delta}}{_{3\gamma}}{}	\;,\nonumber\\
&&m^{(3)}_{1\alpha,2\beta,3\gamma}= \fdif{m^{(2)}_{1\alpha,2\beta}}{\kappa_{3\gamma}}
\equiv\tria{m^{(3)}}{_{1\alpha}}{_{2\beta}}{_{3\gamma}}{}	\;.
\eee
In \mref{5.3} the threepoles are introduced which allows us to write
\mref{5.2} in more convenient form:
\bee
&&\tria{\dot{U}}{_1}{_1}{_3}{}\upp{\hat{m}^{(2)}_{12}}+
\upp{\hat{U}_{11}}\tria{m^{(3)}}{_1}{_2}{_3}{}=\nonumber\\
&&=\sum_{r\in\pi_1}\left(\tria{\dot{G}}{_1}{_r}{_3}{}\upp{\hat{m}^{(2)}_{r2}}+
\upp{\hat{G}_{1r}}\tria{m^{(3)}}{_r}{_2}{_3}{}\right)
\label{(4.4)}	\;.
\eee
It should be noted that equality of site indices in threepoles and matrices
implies
also the symmetry with respect to corresponding sort indices, \ie
\be
\tria{\dot{U}}{_{1\gamma}}{_{1\delta}}{_{3\epsilon}}{}=\tria{\dot{U}}{_{1\delta}}{_{1\gamma}}{_{3\epsilon}}{}
\;.
\ee
Differentiating expressions \mref{4.17} and \mref{4.19} we find
\bee
&&\tria{\dot{U}}{_1}{_1}{_3}{}=
z'\upp{\ptsed{\hat{F}_1^{(2)}}^{-1}}\tria{\dot{F}^{(2)}}{_1}{_1}{_3}{}\upp{\ptsed{\hat{F}_1^{(2)}}^{-1}}+\nonumber\\
&&+\sum_{r\in\pi_1}\upp{\hat{E}_{11}}\left(\tria{\dot{F}^{(11)}}{_1}{_r}{_3}{}\upp{\ptsed{\hat{F}_{r1}^{(20)}}^{-1}\hat{F}_{r1}^{(11)}}-\right.\nonumber\\
&&-\upp{\hat{F}_{1r}^{(11)}\ptsed{\hat{F}_{r1}^{(20)}}^{-1}}\tria{\dot{F}^{(20)}}{_r}{_r}{_3}{}\upp{\ptsed{\hat{F}_{r1}^{(20)}}^{-1}\hat{F}_{r1}^{(11)}}+\nonumber\\
&&\left.+\upp{\hat{F}_{1r}^{(11)}\ptsed{\hat{F}_{r1}^{(20)}}^{-1}}\tria{\dot{F}^{(11)}}{_r}{_1}{_3}{}-\tria{\dot{F}^{(20)}}{_1}{_1}{_3}{}\right)\upp{\hat{E}_{11}}
\label{4.5}	\;.
\eee
When obtaining the formula \mref{4.5} the identities of the following
type are used:
\be
\fdif{\ptsed{\hat{F}^{(2)}_1}^{-1}}{\kappa_{3\gamma}}=-\ptsed{\hat{F}^{(2)}_1}^{-1}
\ptsed{\fdif{\hat{F}^{(2)}_1}{\kappa_{3\gamma}}}\ptsed{\hat{F}^{(2)}_1}^{-1}
\;.
\ee
They can be obtained for any matrix $\hat{A}$ by means of differentiation
of the identity $\hat{A}\hat{A}^{-1}=1$.

	Now we shall obtain the expressions for the threepoles which have
appeared in \mref{4.5}
\be
\fdif{(\hat{F}^{(2)}_1)\ab}{\kappa_{3\gamma}}=
\fdif{\avX{F^{(2)}_{1\alpha\beta}}}{\kappa_{3\gamma}}=
\avX{F^{(3)}_{1\alpha\beta\delta}}\hat{\kappat}^{(1)}_{1\delta,3\gamma}
\;.
\ee
In the diagram form it can be represented as
\be
\tria{\dot{F}^{(2)}}{_1}{_1}{_3}{}=\tria{F^{(3)}}{_1}{_1}{_1}{\hat{\kappat}^{(1)}_{13}}
\label{4.7}	\;.
\ee
The other threepoles we find in the same way:
\be
\tria{\dot{F}^{(20)}}{_r}{_r}{_3}{}=
\tria{F^{(30)}}{_r}{_r}{_r}{\,_1\hat{\kappat}^{(1)}_{r3}}+
\tria{F^{(21)}}{_r}{_r}{_1}{\,_r\hat{\kappat}^{(1)}_{13}}
\label{4.9}	\;,
\ee
\be
\tria{\dot{F}^{(11)}}{_1}{_r}{_3}{}=
\tria{F^{(21)}}{_1}{_r}{_1}{\,_r\hat{\kappat}^{(1)}_{13}}+
\tria{F^{(21)}}{_1}{_r}{_r}{\,_1\hat{\kappat}^{(1)}_{r3}}	\;,
\ee
where
\bee
&&\tria{F^{(30)}}{_{1\alpha}}{_{1\beta}}{_{1\gamma}}{}=
\avX{F^{(30)}_{1\alpha\beta\gamma,r}}=
\avX{\ang{S_{1\alpha}S_{1\beta}S_{1\gamma}}^c_{\rho_{1r}} }	\;,\nonumber\\
&&\tria{F^{(21)}}{_{r\alpha}}{_{r\beta}}{_{1\gamma}}{}=
\avX{F^{(21)}_{r\alpha\beta,1\gamma}}=
\avX{ \ang{S_{r\alpha}S_{r\beta}S_{1\gamma}}^c_{\rho_{1r}} }	\;,\nonumber\\
&&\tria{F^{(21)}}{_{1\alpha}}{_{r\beta}}{_{1\gamma}}{}=
\avX{ F^{(21)}_{1\alpha\gamma,r\beta} }=
\avX{ \ang{S_{1\alpha}S_{1\gamma}S_{r\beta}}^c_{\rho_{1r}} }	\;,\nonumber\\
&&\tria{F^{(12)}}{_{1\alpha}}{_{r\beta}}{_{r\gamma}}{}=
\avX{ F^{(12)}_{1\alpha,r\beta\gamma} }=
\avX{ \ang{S_{1\alpha}S_{r\beta}S_{r\gamma}}^c_{\rho_{1r}} }	\;.
\eee
The quantity $_r\hat{\kappat}^{(1)}_{13}$ which appears in \mref{4.9} follows
from \mref{kf}, \mref{4.16} and can be expressed as
\be
_r\hat{\kappat}^{(1)}_{13}=\hat{\kappat}^{(1)}_{13}-\,_r\hat{\bar{\varphi}}^{(1)}_{13}=
-\hat{E}_{11}\hat{m}^{(2)}_{13}-\hat{G}_{1r}\hat{m}^{(2)}_{r3}
\label{(4.11)}	\;.
\ee
Using \mref{4.7}-\mref{4.9} in \mref{4.5} and taking into account
\mref{(4.11)}, \mref{mefk} we get the expression for the threepole
$\dot{U}$ involving only
the quantities calculated above, \ie the matrices and threepoles
$F$ and the pair CF $\hat{m}^{(2)}_{ij}$. The same procedure is used to
calculate the threepole $\dot{G}$. Using the obtained expressions in
\mref{(4.4)} we get the equation for the 3-site CF:
\bee
&&\upp{\hat{U}_{11}}\tria{m^{(3)}}{_1}{_2}{_3}{}=
\sum_{r\in\pi_1}\left(\tria{A}{_1}{_1}{_1}{\hat{m}^{(2)}_{13}}
\upp{\hat{m}^{(2)}_{12}}+\right.\nonumber\\
&&\left.\tria{B}{_1}{_1}{_r}{\hat{m}^{(2)}_{r3}}\upp{\hat{m}^{(2)}_{12}}+
\tria{C}{_1}{_r}{_1}{\hat{m}^{(2)}_{13}}\upp{\hat{m}^{(2)}_{r2}}+
\tria{D}{_1}{_r}{_r}{\hat{m}^{(2)}_{r3}}\upp{\hat{m}^{(2)}_{r2}}\right)+\nonumber\\
&&+\sum_{r\in\pi_1}\upp{\hat{G}_{1r}}\tria{m^{(3)}}{_r}{_2}{_3}{}
\label{4.12}	\;,
\eee
where
\bee
&&\tria{A}{_1}{_1}{_1}{}=\frac{z'}{z}\upp{(\hat{F}^{(2)}_1)^{-1}}
\tria{F^{(3)}}{_1}{_1}{_1}{(\hat{F}^{(2)}_1)^{-1}}\upp{(\hat{F}^{(2)}_1)^{-1}}-\nonumber\\
&&-\left(\upp{\hat{G}_{1r}}\tria{F^{(30)}}{_r}{_r}{_r}{\hat{G}_{r1}}\upp{\hat{G}_{r1}}+
\upp{\hat{G}_{1r}}\tria{F^{(21)}}{_r}{_r}{_1}{\hat{E}_{11}}\upp{\hat{G}_{r1}}+\right.\nonumber\\
&&+\left.\upp{\hat{E}_{11}}\tria{F^{(21)}}{_1}{_r}{_r}{\hat{G}_{r1}}\upp{\hat{G}_{r1}}+
\upp{\hat{G}_{1r}}\tria{F^{(21)}}{_r}{_1}{_r}{\hat{G}_{r1}}\upp{\hat{E}_{11}}\right)-\nonumber\\
&&-\left\{\upp{\hat{E}_{11}}\tria{F^{(30)}}{_1}{_1}{_1}{\hat{E}_{11}}\upp{\hat{E}_{11}}+
\upp{\hat{E}_{11}}\tria{F^{(21)}}{_1}{_1}{_r}{\hat{G}_{r1}}\upp{\hat{E}_{11}}+\right.\nonumber\\
&&+\left.\upp{\hat{G}_{1r}}\tria{F^{(21)}}{_r}{_1}{_1}{\hat{E}_{11}}\upp{\hat{E}_{11}}+
\upp{\hat{E}_{11}}\tria{F^{(21)}}{_r}{_r}{_1}{\hat{E}_{11}}\upp{\hat{G}_{r1}}\right\}
\label{3poles:a}	\;,\\
&&\tria{B}{_1}{_1}{_r}{}=
-\left(\upp{\hat{G}_{1r}}\tria{F^{(30)}}{_r}{_r}{_r}{\hat{E}_{rr}}\upp{\hat{G}_{r1}}+
\upp{\hat{G}_{1r}}\tria{F^{(21)}}{_r}{_r}{_1}{\hat{G}_{1r}}\upp{\hat{G}_{r1}}+\right.\nonumber\\
&&+\left.\upp{\hat{E}_{11}}\tria{F^{(21)}}{_1}{_r}{_r}{\hat{E}_{rr}}\upp{\hat{G}_{r1}}+
\upp{\hat{G}_{1r}}\tria{F^{(21)}}{_r}{_1}{_r}{\hat{E}_{rr}}\upp{\hat{E}_{11}}\right)-t.i.
\label{3poles:b}	\;,\\
&&\tria{C}{_1}{_r}{_1}{}=
-\left(\upp{\hat{G}_{1r}}\tria{F^{(30)}}{_r}{_r}{_r}{\hat{G}_{r1}}\upp{\hat{E}_{rr}}+
\upp{\hat{G}_{1r}}\tria{F^{(21)}}{_r}{_r}{_1}{\hat{E}_{11}}\upp{\hat{E}_{rr}}+\right.\nonumber\\
&&+\left.\upp{\hat{E}_{11}}\tria{F^{(21)}}{_1}{_r}{_r}{\hat{G}_{r1}}\upp{\hat{E}_{rr}}+
\upp{\hat{G}_{1r}}\tria{F^{(21)}}{_r}{_1}{_r}{\hat{G}_{r1}}\upp{\hat{G}_{1r}}\right)-t.i.
\label{3poles:c}	\;,\\
&&\tria{D}{_1}{_r}{_r}{}=
-\left(\upp{\hat{G}_{1r}}\tria{F^{(30)}}{_r}{_r}{_r}{\hat{E}_{rr}}\upp{\hat{E}_{rr}}+
\upp{\hat{G}_{1r}}\tria{F^{(21)}}{_r}{_r}{_1}{\hat{G}_{1r}}\upp{\hat{E}_{rr}}+\right.\nonumber\\
&&+\left.\upp{\hat{E}_{11}}\tria{F^{(21)}}{_1}{_r}{_r}{\hat{E}_{rr}}\upp{\hat{E}_{rr}}+
\upp{\hat{G}_{1r}}\tria{F^{(21)}}{_r}{_1}{_r}{\hat{E}_{rr}}\upp{\hat{G}_{1r}}\right)-t.i.
\label{3poles:d}
\eee
In \mref{3poles:a}-\mref{3poles:d} "t.i." means twin to the previous
(in parentheses) item, \ie item where the matrix substitution
$\hat{E}\leftrightarrow \hat{G}$
has been carried out. The last (in braces) item of \mref{3poles:a} is an example of
t.i. to the previous (in parentheses) one.

	Assuming uniform field and performing the Fourier transformation
\bee
&&\tria{m^{(3)}}{_i}{_j}{_k}{}=
\tria{m^{(3)}}{}{}{}{}(\vec{r}_j-\vec{r}_i,\vec{r}_k-\vec{r}_i)=\label{4.14}\\
&&{V_e\over (2\pi)^D}\int d\vec{q}_1\;
e^{-i\vec{q}_1(\vec{r}_j-\vec{r}_i)}
{V_e\over (2\pi)^D}\int d\vec{q}_2\;
e^{-i\vec{q}_2(\vec{r}_k-\vec{r}_i)}
\tria{m^{(3)}}{}{}{}{}(\vec{q}_1,\vec{q}_2)  \nonumber
\eee
\begin{figure}
\unitlength=1mm
\centerline{
 \begin{picture}(120,130)
 \put(0,130){\special{em:graph td.pcx}}
 \end{picture}
}
\caption{
Temperature dependencies of spin-spin and
sort-sort CFs' $\vec{q}=0$ Fourier-transforms and of specyfic heat
(in units of $k_B$) of pure and dilute ($c_1=0.6$) systems on plane
square lattice ($z=4$).
}
\end{figure}
we can obtain from \mref{4.12} the following expression for the 3-site CF
\bee
&&\tria{m^{(3)}}{}{}{}{}(\vec{q}_1,\vec{q}_2)=\upp{\hat{m}^{(2)}(\vec{q}_1+\vec{q}_2)}
\left( z\tria{A}{}{}{}{}+\right.\nonumber\\
&&\pi(\vec{q}_1)\tria{B}{}{}{}{}+\pi(\vec{q}_2)\tria{C}{}{}{}{}+\nonumber\\
&&\left.\pi(\vec{q}_1+\vec{q}_2)\tria{D}{}{}{}{}\right)
\unitlength=\trialen
\begin{picture}(10,8)
\put(0,9){\makebox(0,0)[lc]{$\hat{m}^{(2)}(\vec{q}_1)$}}
\put(0,-8){\makebox(0,0)[lc]{$\hat{m}^{(2)}(\vec{q}_2)$}}
\end{picture}	\qquad,
\eee
where the equality which follows from \mref{OZeq}
\be
\hat{U}-\pi (\vec{q}_1+\vec{q}_2)\hat{G}=\left(\hat{m}^{(2)}(\vec{q}_1+\vec{q}_2)\right)^{-1}
\ee
has been taken into account.

\sect{Annealed case}
\newcommand{\hinterval}{\hfill}
\newcommand{\vinterval}{\\[3ex]}
\newcommand{\xlabel}{{\footnotesize$qa$}}
\newcommand{\ylabel}{{\footnotesize$m^{(2)}(\vec{q})$}}
\newcommand{\narrowline}{\special{em:linewidth 0.1pt}}
\newcommand{\thinline}{\special{em:linewidth 0.5pt}}
\newcommand{\putlabels}{
\put(220,920){\makebox(0,0)[lb]{\ylabel}}
\put(1450,90){\makebox(0,0)[lb]{\xlabel}}
\put(1306,812){\makebox(0,0)[r]{TCA}}
\put(1328,812){\special{em:moveto}}
\put(1394,812){\special{em:lineto}}
\put(1306,712){\makebox(0,0)[r]{MFA}}
\multiput(1328,712)(20.756,0.000){4}{\usebox{\plotpoint}}
\put(1394,712){\usebox{\plotpoint}}
}
\setlength{\unitlength}{0.11pt}
In this case we shall introduce into Hamiltonian additional variation
parameters $\bar{\psi}$. The quantity $\ps\ia$
plays the role of effective field acting on the spin of sort $\alpha$ in the
site $i$ from its neighbour (of any sort) in the site $r$.
$\Ham$ takes the form
\be
\Ham = \sum_i H_i + \sum_{(ij)}W_{ij}	\;,
\ee
where
\bee
&&H_i = \kappat_i S_i+\bar{\mu}_i ;
   \quad \bar{\mu}_i = \mu_i + \sum_{r\in\pi_i} \ps_i;
   \quad    \ps_i = \sum\as X\ia \ps\ia 	\nonumber\\
&&W_{ij} = -\ _j\bar{\varphi}_i S_i -\ _i\bar{\varphi}_j S_j + K_{ij} S_i S_j
    -\ _j\bar{\psi}_i -\ _i\bar{\psi}_j + V_{ij}	\;.
\eee
Further, the technique of the previous section leads to the generating
function beeing (within TCA) of the following form:
\be
\F(\braced{\kappa,\mu})=-z'\sum_i\F_i+\sum_{(ij)}\F_{ij}	\;,
\ee
where 
\be
\F_{ij}=\ln \Z_{ij};\qquad \Z_{ij}=\Sp_{S_iS_j} e^{H_{ij}}	\;;\\
\ee
\bee
H_{ij}&=&\,_j\kappat_i S_i+\,_i\kappat_j S_j+K_{ij} S_i S_j + \nonumber\\
&+&\,_j\bar{\mu}_i+\,_i\bar{\mu}_j+V_{ij}	\;;
\eee
\be
\,_j\bar{\mu}_i=\bar{\mu}_i-\,_j\bar{\psi}_i=
\mu_i+\sum_{r\in\pi_i,r\ne j}\ps_i	\;.
\ee
The conditions
\be
\pdif{\F}{\fr\ia} = 0 ; \qquad \pdif{\F}{\ps\ia} = 0
\ee
lead to
\bee
&&\ang{S\ia}=\ang{S\ia}_{\rho_i}=\ang{S\ia}_{\rho_{ij}}\label{4.3a}	\;,\\
&&\ang{X\ia}=\ang{X\ia}_{\rho_i}=\ang{X\ia}_{\rho_{ij}}\label{4.3b}	\;,
\eee
where
\be
\rho_i=\Z_i^{-1}\exp H_i;\qquad \rho_{ij}=\Z_{ij}^{-1}\exp H_{ij};
\qquad j\in\pi_i	\;,
\ee
\ie again equality of the one-site and intracluster unary CFs is implied.
We note that equations \mref{4.3b} like the ones for the chemical potential
\mref{mu equation} are not independent due to the identity
\be
\sum_\alpha\ang{X\ia}=1 \label{4.4}	\;.
\ee
Therefore the number of independent fields $\ps\ia$ and chemical potentials
$\mu\ia$ decreases. We put
\be
\ps_{i\Omega}\equiv 0;\qquad\mu_{i\Omega}\equiv 0 \label{a4.5}	\;.
\ee

In order to shorten following formulae let us intoduce a notation
$\largeS=\braced{S,X}$, \ie
\[
 \largeS\ia = \left\{
\begin{array}{lc}
        S\ia, & $if $ \alpha\le\Omega\\
        X_{i\alpha-\Omega}, & $if $ \alpha>\Omega
\end{array}
             \right.	,
\]
and $\largeKappa=\braced{\kappa,\mu}$, $\largePhi=\braced{\varphi,\psi}$,
\be
M^{(n)}_{i_1\alpha_1\cdots i_n\alpha_n}=\ang{
\largeS_{i_1\alpha_1}\cdots \largeS_{i_n\alpha_n}
}^c=
\fdif{}{\largeKappa_{i_1\alpha_1}}\cdots\fdif{}{\largeKappa_{i_n\alpha_n}}
\F(\braced{\largeKappa})
\ee
With this new notations \mref{4.3a} and \mref{4.3b} can be written as
\be
M^{(1)}\ia=\ang{\largeS\ia}_{\rho_i}=
\ang{\largeS\ia}_{\rho_{ij}}\label{4.3ab}	\;,
\ee

The same technique as for quenched case yields a result for a pair CF
in the same form:
\bee
 \ptsed{ \hat{M}^{(2)}(\vec{q}) }^{-1} &=& (1-z)\ptsed{\hat{\F}^{(2)}}^{-1}+
        z\ptsed{\hat{\F}^{(20)}+\hat{\F}^{(11)}}^{-1}+\nonumber\\
   &+&  \ptsed{\pi(\vec{0})-\pi(\vec{q})}
   \ptsed{\hat{\F}^{(20)}\ptsed{\hat{\F}^{(11)}}^{-1}\hat{\F}^{(20)}
  - \hat{\F}^{(11)} }^{-1}         \label{4.6}	\;,
\eee
where
\be
\ptsed{\hat{M}^{(2)}_{ij}}\ab =
\fdif{^2\F}{\largeKappa\ia\partial\largeKappa\jb}
=\ang{\largeS\ia\largeS\jb}^c
=\pmatrix{\ang{\hat{S_iS_j}}^c&\ang{\hat{S_iX_j}}^c\cr
\ang{\hat{X_iS_j}}^c&\ang{\hat{X_iX_j}}^c}	\;;
\ee\be
\ptsed{\hat{\F}^{(2)}}\ab=\ang{\largeS\ia\largeS_{i\beta}}^c_{\rho_i};\;
\ptsed{\hat{\F}^{(20)}}\ab=\ang{\largeS\ia\largeS_{i\beta}}^c_{\rho_{ij}};\;
\ptsed{\hat{\F}^{(11)}}\ab=\ang{\largeS\ia\largeS\jb}^c_{\rho_{ij}}
\label{intraclu:annealed}	\;.
\ee
In equation \mref{4.6} the sort indices lay in the restricted interval
$\alpha=1\cdots 2\Omega-1$ due to \mref{4.4}, \mref{a4.5};
and the matrices \mref{intraclu:annealed} have the dimension
$(2\Omega-1)\times(2\Omega-1)$. The exact result \cite{Balagurov}
for the one-dimensional system ($z=2$) is a partial case of the formula
\mref{intraclu:annealed}.

The relation \mref{4.3ab} is equivalent to
\be
\Sp_{S_j,X_j} \rho_{ij}=\rho_i \;,
\ee
and it yields
\be
\F^{(n0)}_{ij}=\F^{(n)}_i	\;.
\ee
\renewcommand{\putlabels}{
\put(220,920){\makebox(0,0)[lb]{\ylabel}}
\put(1450,90){\makebox(0,0)[lb]{\xlabel}}
\put(1306,812){\makebox(0,0)[r]{quenched}}
\put(1328,812){\special{em:moveto}}
\put(1394,812){\special{em:lineto}}
\put(1306,712){\makebox(0,0)[r]{MFA}}
\multiput(1328,712)(20.756,0.000){4}{\usebox{\plotpoint}}
\put(1394,712){\usebox{\plotpoint}}
\thinline
\put(1306,612){\makebox(0,0)[r]{annealed}}
\put(1328,612){\special{em:moveto}}
\put(1394,612){\special{em:lineto}}
\narrowline
}
\begin{figure}
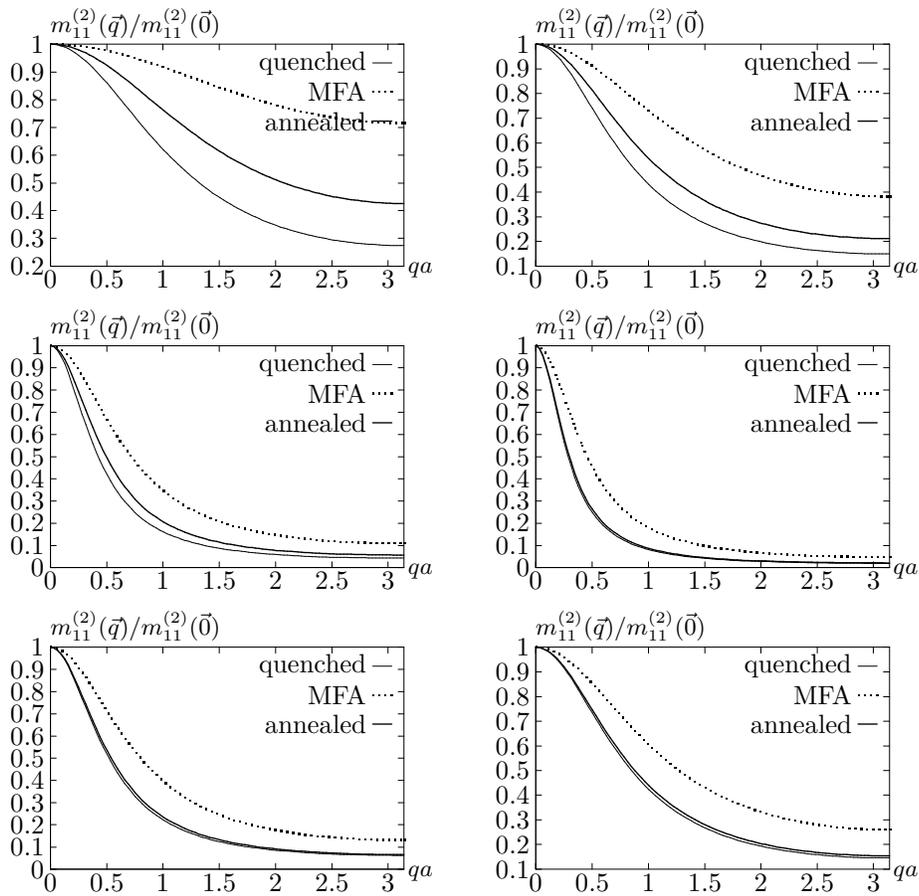

\vspace{2em}
\renewcommand{\ylabel}
{{\footnotesize$m_{11}^{(2)}(\vec{q})/m_{11}^{(2)}(\vec{0})$}}
\input 06qn2.pic
\caption{
$\vec{q}$-dependence of the diluted system CFs. $c_1=0.6$, $V\ab=0$.
$T/T_c=0.5,0.7,0.9,1.1,1.3,1.7$. $q=q_x=q_y$.
}
\label{06qdep}
\end{figure}
Therefore \mref{4.6} can be written in more simple form
\be
 \hat{M}^{(2)}(\vec{q}) = \hat{\F}^{(2)}
\ptsed{1-f^2}^{-1}\ptsed{1+z'f^2-\pi(\vec{q})f};\quad
f=\ptsed{\F^{(2)}}^{-1}\F^{(11)}\;.
\ee

An expression for the 3-site CFs also remains the same form
as in the quenched case
\bee
&&\tria{M^{(3)}}{}{}{}{}(\vec{q}_1,\vec{q}_2)=\upp{\hat{M}^{(2)}(\vec{q}_1+\vec{q}_2)}
\left( z\tria{A}{}{}{}{}+\right.\nonumber\\
&&\pi(\vec{q}_1)\tria{B}{}{}{}{}+\pi(\vec{q}_2)\tria{C}{}{}{}{}+\nonumber\\
&&\left.\pi(\vec{q}_1+\vec{q}_2)\tria{D}{}{}{}{}\right)
\unitlength=\trialen
\begin{picture}(10,8)
\put(0,9){\makebox(0,0)[lc]{$\hat{M}^{(2)}(\vec{q}_1)$}}
\put(0,-8){\makebox(0,0)[lc]{$\hat{M}^{(2)}(\vec{q}_2)$}}
\end{picture}	\qquad,
\eee
with $A$, $B$, $C$, $D$ given by \mref{3poles:a}-\mref{3poles:d} and
intracluster CFs 
\bee
&&\tria{\F^{(30)}}{_{i\alpha}}{_{i\beta}}{_{i\gamma}}{}=
\pdif{^3\F}{\largeKappa\ia\partial\largeKappa_{i\beta}\partial\largeKappa_{i\gamma}}=
\ang{\largeS_{i\alpha}\largeS_{i\beta}\largeS_{i\gamma}}^c_{\rho_{ij}}	\;,\nonumber\\
&&\tria{\F^{(21)}}{_{j\alpha}}{_{j\beta}}{_{i\gamma}}{}=
\pdif{^3\F}{\largeKappa_{j\alpha}\partial\largeKappa\jb\partial\largeKappa_{i\gamma}}=
\ang{\largeS_{j\alpha}\largeS_{j\beta}\largeS_{i\gamma}}^c_{\rho_{ij}}	\;,\nonumber\\
&&\tria{\F^{(21)}}{_{i\alpha}}{_{i\beta}}{_{j\gamma}}{}=
\pdif{^3\F}{\largeKappa\ia\partial\largeKappa_{i\beta}\partial\largeKappa_{j\gamma}}=
\ang{\largeS_{i\alpha}\largeS_{i\beta}\largeS_{j\gamma}}^c_{\rho_{ij}}	\;,\nonumber\\
&&\tria{\F^{(12)}}{_{i\alpha}}{_{j\beta}}{_{j\gamma}}{}=
\pdif{^3\F}{\largeKappa\ia\partial\largeKappa_{j\beta}\partial\largeKappa_{j\gamma}}=
\ang{\largeS_{i\alpha}\largeS_{j\beta}\largeS_{j\gamma}}^c_{\rho_{ij}}	\;.
\eee
\sect{Discussion}
  The great fluctuations of quantities under discussion make the
results of "effective field" theories (like TCA and MFA) worse.
Thus for the present model a great difference in interactions
(e.g., $K\ab=K\delta_{1\alpha}$ (sort 1 is diluted by noninteracting
impurties)) is expected to worse the quality of TCA results whereas great
$z$, when interaction $K\iajb$ couples many sites, improves these results.
Taking into account that TCA gives exact results for the
one-dimensional system ($z=2$), the diluted system on plane square
lattice seems to be the most difficult test for TCA, which can
discover all its shortcomings. That is why below we concentrate
our attention at this case.

  Now let us consider a numerical investigation of the
obtained results. In the quenched case we assume for simplicity
$w\ab=c\as c_\beta$ (complete chaos: $\avX{X\ia X\jb}=\avX{X\ia}\avX{X\jb}$).
The figures \ref{PhaseDiags1}, \ref{PhaseDiags2} show model's phase
diagrams of two-sort system for
different parameters of the Hamiltonian. Curie temperature $T_c$ and
temperature of alloy spinodal segregation $T_s$ are found from the condition
$M\ab^{(2)}(\vec{q}=0,T)\rightarrow \infty$.
Apart from a complicated form of the TCA phase diagrams
\ref{PhaseDiags1} with respect to those of MFA \ref{PhaseDiags2}
one can also remark the rule, that great values of the quantities
$\tilde{K}=K_{11}+K_{22}-2K_{12}$, $\tilde{J}=J_{11}+J_{22}-2J_{12}$
enhance both segregation and spin alignment, whereas great
$\tilde{V}_{11}=V_{11}+V_{22}-2V_{12}$,
$\tilde{I}_{11}=I_{11}+I_{22}-2I_{12}$ enhance segregation and therefore
enhance spin alignment, if $\tilde{K}>0$, $\tilde{J}>0$, and supress
alignment, if $\tilde{K}<0$, $\tilde{J}<0$.

Temperature dependence of order parameters $m_1^{(1)}/c_1$ or
$M_1^{(1)}/c_1$ of pure and dilute systems are shown in the
fig. \ref{diferent M}. TCA and MFA results are compared.
The known exact result for the
pure system on the plane square lattice is also depicted.
For quenched system within TCA the quantity $\sigma^{(1)}=m^{(1)}_1/c_1$
stays less than $1$ even at $T=0$ and vanishes at $c_1<c_p=\frac 1{z-1}$
(below percolation point).
This contradicts to MFA result $\sigma^{(1)}=1$ at $T=0$ for
all $c_1$. Moreover, within MFA $\sigma^{(1)}$ and
$\sigma^{(2)}=J_{11}m^{(2)}_{11}(\vec{q})/(c_1T)$ depend on temperature,
concentration and interaction strength only via $T/T_c^{MFA}$
($T_c^{MFA}=c_1J_{11}/k_B$): $\sigma^{(1)}=\sigma^{(1)}(T/T_c^{MFA})$,
$\sigma^{(2)}=\sigma^{(2)}(\vec{q},T/T_c^{MFA})$. In TCA this takes place
for the case $w_{11}=c_1$ only (all interacting spins constitute one infinite
cluster).

One can see in the figure \ref{m(q)} that within TCA pair CF of the
\begin{figure}
\unitlength=1mm
\centerline{
 \begin{picture}(120,190)
 \put(0,190){\special{em:graph m_qandt.pcx}}
 \end{picture}
}
\caption{
CFs of the ideal ($c_1=1$) and diluted ($c_1=0.6$) systems.
}
\label{m(q)}
\end{figure}
quenched diluted system
$m^{(2)}_{11}(\vec{q})$ is not zero at $T=0$ and it yields infinite rise
of the susceptibility at zero temperature
$\chi(\vec{q})\stackrel{T\rightarrow 0}{\longrightarrow} \infty$
($\chi(\vec{q})\sim m_{11}^{(2)}(\vec{q})/T$) 
whereas MFA gives $\chi(\vec{q})\rightarrow 0$
similarly to the case of the pure
system. The TCA behavior of $\chi(\vec{q})$ and $\sigma^{(1)}$
can be attributed to interacting spins out of the infinite cluster of
interacting spins (finite size cluster effect \cite{Kaneyoshi}). Indeed,
one can find that susceptibility of the isolated spins, which are
always present in diluted system, diverges at $T=0$. Finite size cluster
effects are observed also in
other approximations (e.g., $\chi(\vec{0})$ of the quenched diluted system
in the effective field approximation of Kaneyoshi et.al. \cite{Kaneyoshi}).
Figure \ref{06qdep} shows, that significant differencies between
annealed and quenched CFs appear at low temperatures.

The ref. \cite{Havlin}, where the pair CF of the pure system
in paraphase within TCA was obtained, reports that $\chi(\vec{q})$
has a maximum as a function of $T$ above $T_c$ for a fixed $\vec{q}$
greater than $\vec{q}_0$. We do not find any maximum for $\chi(\vec{q})$,
but instead we notice a maximum of $m_{11}^{(2)}(\vec{q})$ for all
$\pi(\vec{q})\in \left [ 0,\frac{2zz'}{z'^2+1} \right ]$ in the pure and
diluted systems. It should be
noted that MFA does not find any maximum for $m_{11}^{(2)}(\vec{q})$.

\section*{Appendix}
\setcounter{equation}{0}\renewcommand{\theequation}{A\arabic{equation}}
Here we calculate the thermodynamic potential and CFs of
the system with long-range interaction. Let the Hamiltonian be
\bee
^LH\ptsed{\braced{S}}&=&\sum\ia\Gamma\ia S\ia+ {1\over 2}\sum\iajb K\iajb S\ia S\jb+\nonumber\\
&+&\sum\ia\nu\ia X\ia+ {1\over 2}\sum\iajb V\iajb X\ia X\jb+\Delta\,^LH	\;,
\eee
\be
\Delta\,^LH= {1\over 2}\sum\iajb J\iajb S\ia S\jb+
{1\over 2}\sum\iajb I\iajb X\ia X\jb
\label{A2}	\;,
\ee
where the interactions $J\iajb$, $I\iajb$ are not restricted by the nearest
neighbour sites. Consider first the quenched case. Within MFA the
\mref{A2} is taken in the following form
\be
\Delta\,^LH_{MFA}=\sum\iajb \largeJ\iajb\,^Lm^{(1)}\ia \largeS\jb-
{1\over 2}\sum\iajb \largeJ\iajb\,^Lm^{(1)}\ia\,^Lm^{(1)}\jb	\;,
\ee
where $^Lm^{(1)}\ia=\avX{\ang{\largeS\ia}_{^LH}}$ is the one-site CF of the
system, $\largeJ\iajb=\ptsed{\hat{\largeJ}_{ij}}\ab$,
$\hat{\largeJ}_{ij}=\pmatrix{\hat{J}_{ij}&0\cr0&\hat{I}_{ij}}$. 	
The generating function $^L\avX{F_x}(\braced{\Gamma})$ of the system can be
expressed through the generating function
$\avX{F_x}(\braced{\kappa})$ of the reference system \mref{Hamilt}. 
\be
^L\avX{F_x}(\braced{\Gamma})=\avX{F_x}(\braced{\kappa})-
{1\over 2}\sum\iajb J\iajb\,^Lm^{(1)}\ia\,^Lm^{(1)}\jb	\;,
\ee
where
\be
\kappa\ia=\Gamma\ia+\lambda\ia ;\qquad
\lambda\ia=\sum\jb J\iajb\,^Lm^{(1)}\jb
\label{molecular field}	\;.
\ee
Here $\lambda\ia$ is the mean field caused by the long-range
interaction.
Taking into account the relation
\be
\fdif{}{\Gamma\ia}=\sum\jb\fdif{\kappa\jb}{\Gamma\ia} \fdif{}{\kappa\jb}=
\sum\jb\ptsed{\delta\iajb+
\sum_{k\gamma}J_{j\beta,k\gamma}
\,^Lm^{(2)}_{k\gamma,i\alpha}} \fdif{}{\kappa\jb}
\ee
the CFs of the system
\be
^Lm^{(n)}_{i_1\alpha_1\cdots i_n\alpha_n}=
\fdif{}{\Gamma_{i_1\alpha_1}}\cdots \fdif{}{\Gamma_{i_n\alpha_n}}\,^LF (\braced{\Gamma})
\ee
are found to be
\be
^Lm^{(1)}\ia= m^{(1)}\ia	\;,
\ee\be
\label{A9}^Lm^{(2)}\iajb= m^{(2)}_{i\alpha,k\gamma}
\left(\delta_{k\gamma,j\beta}+J_{k\gamma,l\delta}
\,^Lm^{(2)}_{l\delta,j\beta}\right)	\;,
\ee\bee
\label{A10}^Lm^{(3)}_{i\alpha,j\beta,k\gamma}   &=&   m^{(2)}_{i\alpha,l\delta}J_{l\delta,n\eta}m^{(3)}_{n\eta,j\beta,k\gamma}+\nonumber\\
&+& m^{(3)}_{i\alpha,l\delta,m\epsilon}\ptsed{\delta_{l\delta,j\beta}+J_{l\delta,n\eta}\,^Lm^{(2)}_{n\eta,j\beta}}\times\nonumber\\
&\times& \ptsed{\delta_{m\epsilon,k\gamma}+
J_{m\epsilon,o\lambda}\,^Lm^{(2)}_{o\lambda,k\gamma}}	\;,
\eee
where $m^{(1)}$, $m^{(2)}$, $m^{(3)}$ are CFs of the reference system
\mref{Hamilt} in molecular field \mref{molecular field}. We note that in
\mref{A9}, \mref{A10} the sum over
repeated sort and site indices is implied. Fourier transformation
\mref{3.20}, \mref{4.14} allows
us to write \mref{A9}, \mref{A10} in the form
\be
\label{A11}^Lm^{(2)}\ab(\vec{q})= m^{(2)}_{\alpha\gamma}(\vec{q})
\ptsed{\delta_{\gamma\beta}+J_{\gamma\delta}(\vec{q})
\,^Lm^{(2)}_{\delta\beta}(\vec{q})}	\;,
\ee\bee
\label{A12}^Lm^{(3)}_{\alpha\beta\gamma}(\vec{q}_1,\vec{q}_2)  &=&
m^{(2)}_{\alpha\delta}(\vec{q}_1+\vec{q}_2)
J_{\delta\eta}(\vec{q}_1+\vec{q}_2)
m^{(3)}_{\eta\beta\gamma}(\vec{q}_1+\vec{q}_2)+\nonumber\\
&+& m^{(3)}_{\alpha\delta\epsilon}(\vec{q}_1,\vec{q}_2)
\ptsed{\delta_{\delta\beta}+J_{\delta\eta}(\vec{q}_1)\,^Lm^{(2)}(\vec{q}_1)}\times\nonumber\\
&\times&\ptsed{\delta_{\epsilon\gamma}+J_{\epsilon\lambda}(\vec{q}_2)
\,^Lm^{(2)}_{\lambda\gamma}(\vec{q}_2)}	\;.
\eee
The solutions of equations \mref{A11}, \mref{A12} read
\be
\label{A13}
^L\hat{m}^{(2)}(\vec{q})=\ptsed{1-\hat{m}^{(2)}(\vec{q})
\hat{J}(\vec{q})}^{-1}\hat{m}^{(2)}(\vec{q})	\;,\\
\ee\be
\label{A14}\tria{^Lm^{(3)}}{}{}{}{}(\vec{q}_1,\vec{q}_2)=\upp{\ptsed{1-\hat{m}^{(2)}(\vec{q})\hat{J}(\vec{q})}^{-1}}
\tria{m^{(3)}}{}{\ptsed{1-\hat{J}(\vec{q})\hat{m}^{(2)}(\vec{q})}^{-1}}{}
{\ptsed{1-\hat{J}(\vec{q})\hat{m}^{(2)}(\vec{q})}^{-1}}
(\vec{q}_1,\vec{q}_2)~~~~~~~~~~~~~~.
\ee
In \mref{A14} the equality
\be
1+\hat{J}(\vec{q})\,^L\hat{m}^{(2)}(\vec{q})=
\ptsed{1-\hat{J}(\vec{q})\hat{m}^{(2)}(\vec{q})}^{-1}	\;,
\ee
which follows from \mref{A13} is used.

In the case of annealed system the same scheme leads to similar relations.
\bee
&&^L\F (\braced{\largeGamma})=
\F (\braced{\largeKappa})- {1\over 2}\sum\iajb
\largeJ\iajb	\,^LM^{(1)}\ia	\,^LM^{(1)}\jb	\label{long freeen}	\;,\\
&&^LM^{(1)}\ia=M^{(1)}\ia	\;,
\eee
where $\largeGamma=\braced{\Gamma,\nu}$,
\be
\largeKappa\ia=\largeGamma\ia+\sum\jb \largeJ\iajb\,^LM^{(1)}\jb	\;,
\ee
and sums over sort indices run within interval $1\cdots2\Omega$.
Taking into account
$\ang{X_{i\Omega}}\equiv 1-\sum_{\alpha=1}^{\Omega-1} \ang{X\ia}$,
\mref{long freeen} can be rewritten in the form
\be
^L\F (\braced{\largeGamma})=
\F (\braced{\largeKappa})- {1\over 2}\sum\iajb
\tilde{\largeJ}\iajb	\,^LM^{(1)}\ia	\,^LM^{(1)}\jb	\;,
\ee
where $1\le \alpha, \beta \le 2\Omega-1$ and
$\hat{\tilde{\largeJ}}_{ij}=\pmatrix{\hat{J_{ij}}&0\cr 0&\hat{\tilde{I}}}$
is a matrix of $(2\Omega-1)\times(2\Omega-1)$ size with
$(\Omega-1)\times(\Omega-1)$ submatrix
$\ptsed{\tilde{I}_{ij}}\ab=I\iajb-I_{i\alpha,j\Omega}-
I_{i\Omega,j\beta}+I_{i\Omega,j\Omega}$. For the two-sort system
($\Omega=2$) we have
\[
\hat{\tilde{\largeJ}}_{ij}=\pmatrix{
J_{i1,j1}&J_{i1,j2}&0\cr
J_{i2,j1}&J_{i2,j2}&0\cr
0&0&I_{i1,j1}-2I_{i1,j2}+I_{i2,j2} }	\;.
\]
Making use of the identity
$\ang{X\ia X_{j\Omega}}^c\equiv
-\sum_{\beta=1}^{\Omega-1} \ang{X\ia X\jb}^c$ one obtains the higher
CFs in the form
\be
^L\hat{M}^{(2)}(\vec{q})=\ptsed{1-\hat{M}^{(2)}(\vec{q})
\hat{\tilde{\largeJ}}(\vec{q})}^{-1}\hat{M}^{(2)}(\vec{q})	\;,
\ee\be
\tria{^LM^{(3)}}{}{}{}{}(\vec{q}_1,\vec{q}_2)=
\upp{\ptsed{1-\hat{M}^{(2)}(\vec{q})\hat{\tilde{\largeJ}}(\vec{q})}^{-1}}
\tria{M^{(3)}}{}{\ptsed{1-\hat{\tilde{\largeJ}}(\vec{q})\hat{M}^{(2)}(\vec{q})}^{-1}}{}
{\ptsed{1-\hat{\tilde{\largeJ}}(\vec{q})\hat{M}^{(2)}(\vec{q})}^{-1}}
(\vec{q}_1,\vec{q}_2)~~~~~~~~~~~~~~~.
\ee

\newpage
\begin{thebibliography}{99}
 \bibitem{Balagurov} B.~Ja.~Balagurov, V.~G.~Vaks, R.~O.~Zajtsev.
Statistics of one-dimensional model of solid solution. // Fiz. Tver. Tela,
1974, vol. 16, No 8, p. 2302-2309. (In Russian).
 \bibitem{Ching} W.~Y.~Ching, D.~L.~Huber. Monte-Carlo studies of the critical
behavior of site-diluted two-dimensional Ising models //Phys. Rev. B,
1976, vol. 13, No 7, p. 2962-2964.
 \bibitem{Stoll} E.~Stoll, T.~Schneider. Ising ferromagnet with quenched
impurities //J. Phys. A, 1976, vol. 9, No 7, p. L67-L70.
 \bibitem{Vaks} V.~G.~Vaks, N.~E.~Zein. On the theory of phase transitions in
solid solution. // J. Exp. Teor. Fiz., 1974, vol. 67, No 9,
p. 1082-1100. (In Russian).
 \bibitem{SatoDilute} H.~Sato, A.~Arrott. Remarks on magnetically dilute
systems. // J. Phys. Chem. Solids, 1959, vol. 10, p. 19-34.
 \bibitem{prepr1} R.~R.~Levitskii, S.~I.~Sorokov, R.~O.~Sokolovskii.
Thermodynamics
and re\-lax\-at\-ion\-al dynamics of disordered Ising model. // Lviv,
1993, 43p. (Preprint / Ukr.Acad.Sci. Inst.Theor.Phys.; IFKS-93-17U).
(In Ukrainian).
 \bibitem{Yukhn} I.~R.~Yukhnovskii, R.~R.~Levytskii, S.~I.~Sorokov.
Thermodynamics and distribution functions of Ising model.
Two-site cluster approximation. // Kiev, 1986, 33p.
(Preprint / Acad. Sci. Ukr.SSR. Inst.Theor.Phys.; ITP-86-142R).
(In Russian).
 \bibitem{Kubo} R.~Kubo. Generalized cumulant expansion method. //J. Phys.
Soc. Japan, 1962, vol. 17, No 7, p. 1100-1120.
\bibitem{Kaneyoshi} T.Kaneyoshi, I.Tamura  and  R.Honmura.  Diluted Ising
ferromagnet: Its physical properties. // Phys.Rev.B, 1984, vol. 29,
p. 2769-2776.
 \bibitem{Havlin} S.~Havlin, H.~Sompolinsky. Cluster approximation for
${\bf q}$-dependent correlations in magnetic and ferroelectric systems.
// Phys. Rev. B, 1982, vol. 25, No 9, p. 5828-5835.
\end{thebibliography}
\end{document}